# Superconductivity in tin telluride films grown by molecular beam epitaxy

Antonio Gonzalez[1], Samuel J. Poage[1], Bernardo Langa, Jr.[2], Deepak Sapkota[2], Salva Salmani-Rezaie[3,4], Shalinee Chikara[5], Michael D. Williams[6], David A. Muller[3,4], Kasra Sardashti[2], and Kaveh Ahadi[1,7,8*]

[1]Department of Materials Science and Engineering, North Carolina State University, Raleigh, NC 27695, USA
[2]Department of Physics and Astronomy, Clemson University, Clemson, SC 29634
[3]School of Applied and Engineering Physics, Cornell University, Ithaca, NY 14853
[4]Kavli Institute at Cornell for Nanoscale Science, Cornell University, Ithaca, NY 14853
[5]National High Magnetic Field Laboratory, Tallahassee, Fl 32310
[6]Department of Physics, Clark Atlanta University, Atlanta, GA 30314
[7]Department of Electrical and Computer Engineering, The Ohio State University, Columbus, OH 43210
[8]Department of Materials Science and Engineering, The Ohio State University, Columbus, OH 43210

[*] Corresponding author.  Email: ahadi.4@osu.edu

**ABSTRACT**

The intersection of superconductivity and ferroelectricity hosts a wide range of exotic quantum phenomena. Here, we report on the observation of superconductivity in high-quality tin telluride films grown by molecular beam epitaxy. Unintentionally doped tin telluride undergoes a ferroelectric transition at ~100 K. The critical temperature of superconductivity increases monotonically with indium concentration. The critical field of superconductivity, however, does not follow the same behavior as critical temperature with indium concentration and exhibits a carrier-density-dependent violation of the Pauli limit. The electron-phonon coupling, from the McMillan formula, exhibits a systematic enhancement with indium concentration, suggesting a potential violation of BCS weak coupling at high indium concentrations.

Topologically nontrivial states in topological crystalline insulators (TCI) are protected by a lattice symmetry, as opposed to $Z_2$ topological insulators in which time reversal symmetry protects these states[1]. Regardless of the nature of topological states, the combination of superconductivity and these states has the potential for topological superconductivity which could enable fault-tolerant quantum computing[2]. Topological superconductivity could emerge intrinsically or be engineered at the interfaces between conventional, s-wave superconductors, and topological insulators[3]. Furthermore, combination of other emergent phenomena and superconductivity could give rise to exotic quantum phenomena[4,5]. For example, a robust superconductivity with enhanced critical temperature is observed when superconductivity emerges from a polar state[6–8]. Spin-orbit coupling (SOC) combined with inversion symmetry breaking elements, i.e., Rashba SOC, in polar superconductors creates helical spin configuration and enhances the superconductivity[9]. Combination of broken inversion symmetry and strong spin-orbit coupling could also give rise to exotic superconducting states including mixed-parity superconductivity[10], topological Weyl superconductivity[11], superconducting diode effect[12], and upper critical field exceeding the Pauli limit[13].

Tin telluride (SnTe) is a narrow bandgap semiconductor (0.3 eV[14,15]) with a cubic rocksalt crystal structure ($Fm\bar{3}m$) at room temperature[16]. It undergoes a ferroelectric transition to a rhombohedral structure ($R3m$) at ~100 K[17–19]. Tin telluride is a topological crystalline insulator in which topological surface states originate from the mirror symmetry with respect to the (110) planes[20]. The existence of the gapless topological surface states was experimentally verified[21,22]. Tin telluride becomes superconducting upon doping[23,24]. Indium-doped bulk samples exhibit critical temperature ($Tc$) as high as ~4.5 K[25]. The solubility limit of indium in bulk tin telluride reaches ~50 at.% at room temperature[26]. The mixed oxidation of indium (+1 and +3) creates amphoteric levels in tin telluride which induces an anomalous dependency to indium concentration[27]. Recent MBE grown layers demonstrated emergence of versatile electronic states ranging from topologically nontrivial states and polar semimetals to superconductors[28,29].

The superconductivity in tin telluride is an ideal platform for the experimental realization of polar superconductivity and nontrivial topology. Here, we report on high-quality molecular beam epitaxy growth and observation of a superconducting transition in indium-doped tin telluride. The critical temperature of superconductivity scales with indium concentration. We also observe the in-plane critical field violating the Pauli paramagnetic limit, depending on the indium density.

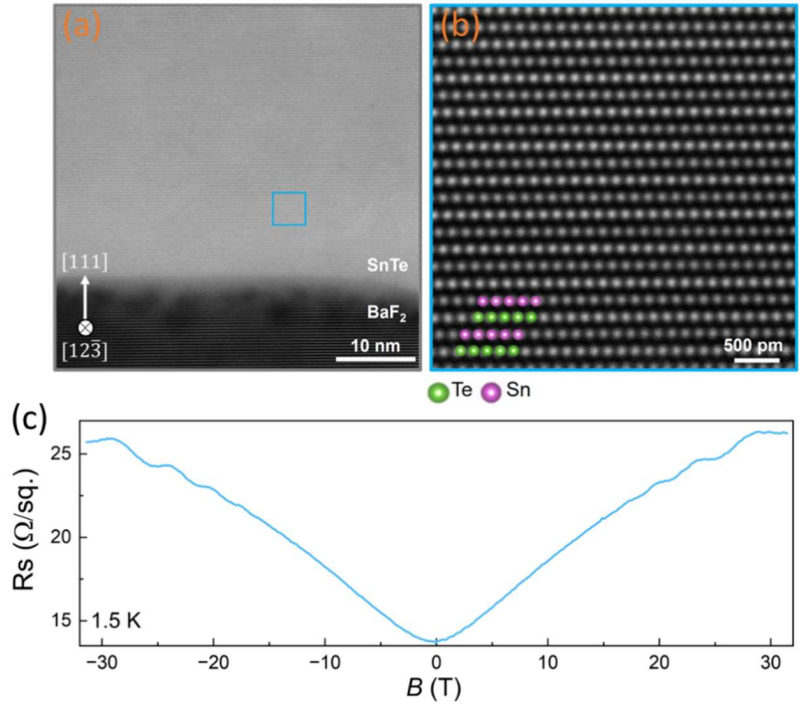

**Figure 1. Molecular beam epitaxy grown tin telluride film on BaF₂ (111).** (a) Cross-sectional HAADF-STEM image of the tin telluride film grown on BaF$_2$ (111). (b) Higher magnification HAADF-STEM image of the tin telluride layer. (c) Longitudinal magneto-resistance measurement of unintentionally doped tin telluride at 1.5 K, exhibiting quantum oscillations.

We used a chalcogenide MBE system (GEN 930) to grow high quality indium-doped tin telluride. The films were grown on BaF$_2$ (111) substrates. BaF$_2$ was chosen due to its minimal lattice and thermal expansion coefficient mismatch with IV-VI materials. BaF$_2$ has a cubic structure ($Fm\bar{3}m$) with lattice constant of ~0.62 nm. The substrate temperature was kept at 400 °C (thermocouple temperature) during the growth. The growth rate was roughly one monolayer per second and the film thickness was kept ~450 nm, corroborated by cross section scanning transmission electron microscopy (STEM) images. We used a tellurium-rich tin telluride source (99.999 at.%, American Elements) to account for tellurium desorption and dopants. The $\theta - 2\theta$ x-ray diffraction (XRD) scan and rocking curve results of the films around 222 reflection can be found in the supplementary materials (Figure S1). The out-of-plane lattice constant ($a_\perp$), extracted from XRD, decreases with increasing indium concentration. The rocking curve full width at half maximum (FWHM) is ~325 arcsecond, suggesting the high crystalline quality of the grown films. We also used STEM to examine the film and interface structure. Cross-sectional specimens were prepared using the Thermo Fisher Scientific Helios G4UX focused ion beam system. High-angle annular dark-field (HAADF) images were captured using the Thermo Fisher Scientific Spectra 300 X-CFEG instrument, running at 300 kV with a 30 mrad convergence angle and an HAADF with an angular range from 60 to 200 mrad. Figure 1(a) exhibits a HAADF-STEM image of SnTe film grown on (111)-oriented $BaF_2$ single crystal substrate. TEM samples were cut along the substrate's edge, which exhibits an 11-degree miscut relative to the primary crystallographic direction of $[1\bar{1}0]$. The magnified section of Figure 1(b) shows alternating layers of tellurium (brighter) and tin (darker) columns. Figure 1(c) shows the magneto-resistance measurement, for the unintentionally doped sample carried out at 1.5 K. The observed quantum oscillations suggest high quality of the grown films. The charge carriers in the unintentionally doped films are due to point defects commonly observed in IV-VI semiconductors[30]. We controlled the indium concentration in the films using indium cell temperature. Secondary ion mass spectroscopy (SIMS) exhibits incorporation of the indium in the films (Figures S2). SIMS also confirms that indium to tin ratio in the grown tin telluride scales with indium cell temperature. The indium concentration profile, however, exhibits an upturn near the surface of films with higher indium concentration. The substantial diffusion of indium and its exothermic substitutional doping, at temperatures as low as room temperature, was previously reported in PbTe [31,32]. Here, we observe indium migration towards the surface, which combined with favorable formation energy of InTe, could explain the indium concentration profile.

Magneto-transport measurements were performed using the Van der Pauw configuration and measuring the differential resistance (i.e., $dV/dI$). The temperature-dependent magneto-transport measurements (except Figure 1(c)) were carried out in a Teslatron refrigerator, Oxford Instruments with a lock-in amplifier in AC mode with excitation current ranging from 1 μA to 10 μA. Measurements below 1.5 K (down to 270 mK) were carried out in the same system using a HelioxVT $^3$He Probe. The magneto-resistance of the unintentionally doped film (Figure 1(c)) was carried out at National High Magnetic Fields Laboratory using a resistive magnet with a lock-in amplifier and an excitation current of 1 μA.

Figure 2 exhibits the normalized resistance with temperature. All samples demonstrate a sharp superconducting transition. The critical temperature of superconductivity, corresponding to 0.5 $R$n, are 2.4 K, 1.3 K, 1.1 K, 0.75 K for indium cell temperatures of 810 °C, 790 °C, 760 °C, and 710 °C, respectively. Figure S3 exhibits the superconducting temperature with respect to indium cell temperature. We estimate the BCS gap ($\triangle_{BCS} \approx 1.76\, K_B T_C$) to be 364 μeV, 194 μeV, 159 μeV, and 139 μeV, for indium cell temperatures of 810 °C, 790 °C, 760 °C, and 710 °C, respectively.

The critical temperature of superconductivity scales with the indium concentration. The SIMS results exhibits the indium concentration reaching as high as ~15 at.% in the sample with 810 °C cell temperature which is well below the solubility limit[26].

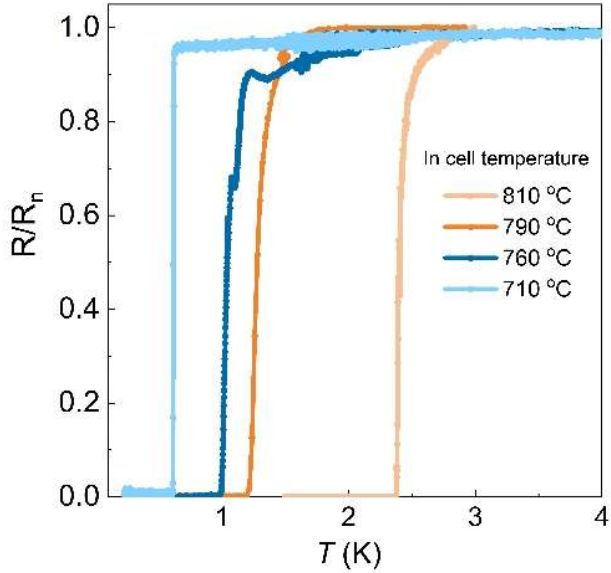

**Figure 2. Critical temperature of superconductivity with indium concentration in tin telluride films.** Longitudinal normalized resistance ($R/R$n, where $R$n is the normal state resistance) of tin telluride films grown on BaF$_2$ (111) with temperature.

Figure 3 shows the superconducting transition with in-plane applied magnetic field at different temperatures. The orbital upper critical field ($H_{C2}$) of superconductivity at 270 mK, corresponding to 0.9 $R$n, are 3.2 T, 2.2 T, 2.2 T, and 1.2 T for indium cell temperatures of 810 °C, 790 °C, 760 °C, and 710 °C, respectively. In the absence of spin-orbit interaction, there are two mechanisms by which applied magnetic field could suppress superconductivity. One is the diamagnetic response related to the action of the field on the orbital motion of electrons forming a Cooper pair. Here, the Ginzburg-Landau coherence length ($\xi_{GL} = \sqrt{\phi_o/2\pi H_{C2}}$, where $\phi_o$ is magnetic quantum flux) using measured orbital upper critical field are 10 nm, 12 nm, 12 nm, and 17 nm, for indium cell temperatures of 810 °C, 790 °C, 760 °C, and 710 °C, respectively, which are well below the thickness of film, suggesting a three dimensional superconductivity. The second mechanism is paramagnetic response associated with the Zeeman splitting of the states with opposite spin, giving rise to Pauli limit ($H_P \approx \Delta/\sqrt{2}\mu_B$, assuming g-factor of 2 and weakly coupled superconductivity). We note, the measured orbital upper critical fields at 270 mK are near the Pauli limit, which we estimate to be, 4.5 T, 2.4 T, 1.9 T, and 1.4 T for indium cell temperatures of 810 ºC, 790 ºC, 760 ºC, and 710 ºC, respectively.

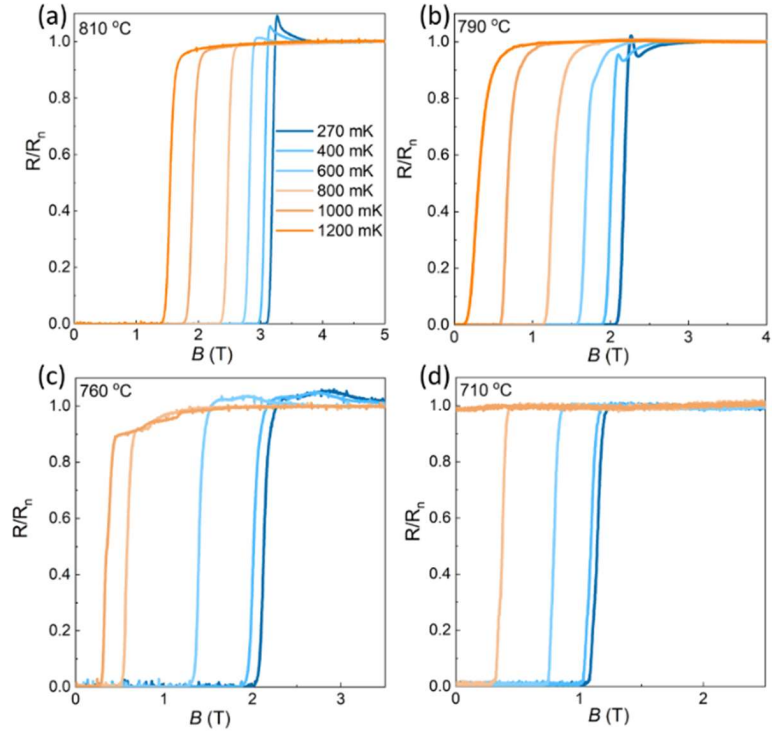

**Figure 3. In-plane critical field of superconductivity with indium concentration in tin telluride films.** Longitudinal normalized resistance ($R/R$n, where $R$n is the normal state resistance) of tin telluride films grown on BaF$_2$ (111) with in-plane magnetic field at various temperatures.

A resistance upturn is observed prior to superconducting transition upon decreasing the field in higher indium concentrations at lower temperatures (Figure 3(a) and (b)). Granular superconductors have exhibited similar behavior, where superconducting puddles form first[33,34]. Furthermore, anisotropic resistance upturn prior to superconducting transition has been reported recently[13]. Here, the resistance upturn could be due to performed Cooper pairs in which gapped out quasiparticles disrupt transport[35] before global superconductivity prevails at lower fields.

Figure 4 shows the in-plane orbital upper critical field ($H_{C2}$) of superconductivity with temperature. The critical field results were fitted using a modified Ginzburg-Landau model ($H_{C2}(T) = H_{C2}(0)(1-t)^\alpha$, where $H_{C2}(0)$ is upper critical field at zero Kelvin and $t = T/T_C$. Temperature dependence of critical field, described by Ginzburg-Landau theory predicts a linear relationship ($\alpha = 1$). The in-plane critical field in a thin film, however, could exhibit a square root characteristic ($\alpha = 0.5$), introduced by Tinkham[36]. Here, the highly doped sample's $H_{C2}$ could be described by a linear fit and lower doped samples exhibit a mixed behavior ($0.5<\alpha<1$) which has been reported previously when the superconductor experiences a dimensional crossover[13]. Here, the calculated coherence length matches previous reports[28,29] and is much smaller than nominal film thickness. The superconducting layer thickness, however, might be challenging to estimate in the films due to indium concentration gradient near the surface (Figure S2).

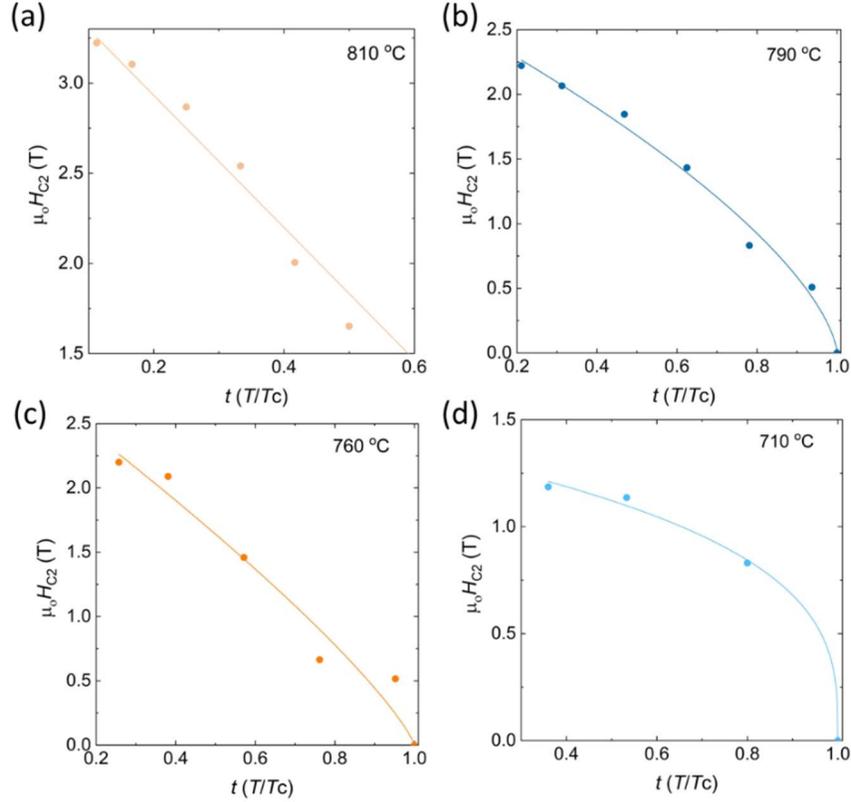

**Figure 4. In-plane critical field of superconductivity with temperature in tin telluride films.** The results are in agreement with a modified Ginzburg-Landau model, $H_{C2}(T) = H_{C2}(0)(1-t)^\alpha$, where $H_{C2}(0)$ is upper critical field at zero Kelvin, $t = T/T_C$ and $\alpha$ is dimensionality factor.

Table 1 summarizes the measured results for tin telluride films with different indium concentrations. The first important conclusion from these results is that the emergence of superconductivity in SnTe depends strongly on the indium concentration. The critical temperature and critical field of superconductivity scale with indium concentration. The in-plane critical field, however, becomes less sensitive to tin concentration around 10% SIMS indium concentration (cell temperatures of 760 °C and 790 °C). We also observe the in-plane critical field reaches the Pauli paramagnetic limit in these films.

**Table I. Summary of superconducting and structural characterization of indium-doped tin telluride films grown by MBE.** Values of lattice constant ($a_\perp$), critical temperature of superconductivity ($T_C$, defined at 0.5Rn), BCS superconducting gap ($\Delta_{BCS} \approx 1.76\, K_B T_C$), in-plane critical field ($H_{C2}$, defined at 0.9Rn), coherence length ($\xi_{GL}$), the critical field to Pauli limit ratio ($H_{C2}(0)/H_P$), and electron-phonon coupling constant ($\lambda_{e-p}$).

| Indium cell temperature | $a_\perp$ (nm) | $T_C$ (K) | $\Delta_{BCS}$ ($\mu eV$) | $H_{C2}$ (T) | $\xi_{GL}$ (nm) | $H_{C2}(0)/H_P$ | $\lambda_{e-p}$ |
|---|---|---|---|---|---|---|---|
| 810 °C | 0.632 | 2.4 | 364 | 3.2 | 10 | 0.8 | 0.63 |
| 790 °C | 0.633 | 1.3 | 194 | 2.2 | 12 | 1.1 | 0.54 |
| 760 °C | 0.633 | 1.1 | 159 | 2.2 | 12 | 1.5 | 0.52 |
| 710 °C | 0.634 | 0.75 | 139 | 1.2 | 17 | 1.0 | 0.49 |

First, we discuss the effect of indium concentration on superconductivity. Insensitivity of the critical field to indium concentrations (760 °C and 790 °C indium cell temperatures) and a change of the critical field behavior with temperature raise the question of whether nature of doping and superconductivity change in tin telluride with indium concentration. Substitution of tin ($Sn^{2+}$) with indium ($In^{1+}$) is expected to induce *p*-type carriers in tin telluride. The self-doped *p*-type (tin-deficient) tin telluride, however, shows different superconducting critical temperatures compared to indium-doped samples with similar carrier densities[24,25,37,38]. Furthermore, an anomalous change in doping character and superconductivity is reported with indium doping in bulk tin telluride at ~10 at.%[27]. The McMillan formula estimates the electron-phonon coupling constant ($\lambda_{e-p}$)[39]. The electron-phonon coupling constant is a dimensionless measure of the strength of electron-phonon coupling. The weak-coupling BCS model is valid for $\lambda_{e-p} \ll 1$.

$$\lambda_{e-p} = \frac{1.04 + \mu^* \ln(\Theta_D/1.45 T_c)}{(1 - 0.62\mu^*) \ln(\Theta_D/1.45 T_c) - 1.04}$$

where $\mu^*$ accounts for the screened Coulomb repulsion. The Debye temperature ($\Theta_D$) was measured for bulk indium-doped tin telluride, ranging from 204 K for 5% indium to 162 K for 40% indium[27]. We used measured critical temperature (*Tc*), approximated the Debye temperature of 200 K for all samples, and made the common assumption of $\mu^*$=0.15[39]. The electron-phonon coupling constant is 0.63, 0.54, 0.52, and 0.49 for indium cell temperatures of 810 °C, 790 °C, 760

°C, and 710 °C, respectively. The films studied here remain within BCS weak coupling regime but the observed enhancement of $\lambda_{e-p}$ with indium suggests potential violation of weak-coupling superconductivity regime in highly doped tin telluride. We note that violation of BCS weak coupling regime was reported in bulk $Sn_{0.6}In_{0.4}Te$[27]. Furthermore, the strongly coupled superconductivity requires a correction to superconducting gap ($\Delta > 1.76\ K_B T_C$)[40] in highly doped samples.

Next, we discuss the in-plane critical field of superconductivity in this system. Here, the $H_{C2}(0)/H_P$ ratio is 0.8, 1.1, 1.5, and 1.0 for indium cell temperatures of 810 °C, 790 °C, 760 °C, and 710 °C, respectively (Figure S3), suggesting the films are near a violation of Pauli limit. In superconducting thin films spin–orbit scattering randomizes the spins and reduces the polarizing effect of the magnetic field[41]. Furthermore, Rashba spin–orbit effect could enhance the critical field up to $\sqrt{2}H_P$[9] due to a distinctive helical spin configuration. The measured $H_{C2}(0)/H_P$ ratio, however, is smaller than other polar superconductors, $SrTiO_3$ (>4)[42] and $KTaO_3$ (>8)[43]. The enhanced critical field of superconductivity in these materials systems is attributed to formation of quasiparticles with extraordinary resilience against magnetic field[43]. The orbital character of charge transport in tin telluride (*p*-orbital), however, is different from $SrTiO_3$ and $KTaO_3$ (*d*-orbital) which could explain the observed discrepancy in $H_{C2}(0)/H_P$ ratio. Here, the spin-orbit scattering and/or Rashba spin-orbit coupling could potentially explain the enhanced critical field.

In summary, our results, especially the dopant density dependence of critical field and nature of superconductivity, should be of interest for theoretical proposals for testing the different models that relate superconductivity to a polar state[44–47]. Independent of the specific mechanism, the results point to opportunities to tune the critical field, the nature of pairing, and searching for new superconducting materials systems that are in proximity to a polar instability. We stress that our findings warrant further study of superconductivity at the intersection of topologically nontrivial states and ferroelectricity.

**Acknowledgments**
SJP was supported by the U.S. National Science Foundation (NSF) under Grant DMR-2408890. KS and BL acknowledge funding from the U.S. National Science Foundation (NSF) under Grant 2137776. This work made use of a Helios FIB supported by the NSF (Grant DMR-1539918) and the Cornell Center for Materials Research (CCMR) Shared Facilities, which are supported through the NSF MRSEC Program (GrantDMR-1719875). MDW also acknowledges support from National Science Foundation PREM Award no. 2122147. A portion of this work was performed at the National High Magnetic Field Laboratory, which is supported by National Science Foundation Cooperative Agreement DMR-2128556 and the State of Florida. BL thanks F. X. Duffy for his assistance in low-temperature experiments. KA acknowledges discussions with M. N. Gastiasoro.

**Data availability**

The data that support the findings of this study are available in the article and its Supporting Information. Raw data can be obtained from the corresponding author upon request.

**Conflict of Interests**

The authors declare that they have no conflict of interest.